\documentclass[twocolumn,epjc3]{svjour3}
\smartqed  
\RequirePackage{graphicx,amssymb,color}
\newcommand{\black}{\color{black}}

\usepackage{graphicx}
\usepackage{mathptmx}      
\usepackage{flushend}
\usepackage[numbers,sort&compress]{natbib}
\usepackage[colorlinks,citecolor=blue,urlcolor=blue,linkcolor=blue]{hyperref}
\usepackage{epstopdf}
\usepackage{caption}
\usepackage{blindtext}
\usepackage[T1]{fontenc}
\usepackage[latin9]{inputenc}
\usepackage{color}
\usepackage{amsmath}
\usepackage{amssymb}
\usepackage{stackrel}
\usepackage{graphicx}
\usepackage{babel}
\usepackage{subfig} 
\usepackage{diagbox}
\usepackage{array}
\usepackage{cuted}
\journalname{Eur. Phys. J. C}
\begin{document}
	
	\global\long\def\a{\alpha}%
\global\long\def\b{\beta}%
\global\long\def\c{\chi}%
\global\long\def\d{\delta}%
\global\long\def\e{\epsilon}%
\global\long\def\f{\phi}%
\global\long\def\g{\gamma}%
\global\long\def\h{\eta}%
\global\long\def\i{\iota}%
\global\long\def\k{\kappa}%
\global\long\def\l{\lambda}%
\global\long\def\m{\mu}%
\global\long\def\n{\nu}%
\global\long\def\o{\omega}%
\global\long\def\p{\pi}%
\global\long\def\q{\theta}%
\global\long\def\r{\rho}%
\global\long\def\s{\sigma}%
\global\long\def\t{\tau}%
\global\long\def\u{\upsilon}%
\global\long\def\x{\xi}%
\global\long\def\y{\psi}%
\global\long\def\z{\zeta}%

\global\long\def\ve{\varepsilon}%
\global\long\def\vf{\varphi}%
\global\long\def\vs{\varsigma}%
\global\long\def\vq{\vartheta}%

\global\long\def\D{\Delta}%
\global\long\def\F{\Phi}%
\global\long\def\G{\Gamma}%
\global\long\def\L{\Lambda}%
\global\long\def\Q{\Theta}%
\global\long\def\S{\Sigma}%
\global\long\def\U{\Upsilon}%
\global\long\def\W{\Omega}%
\global\long\def\X{\Xi}%
\global\long\def\Y{\Psi}%

\global\long\def\6{\partial}%
\global\long\def\8{\infty}%
\global\long\def\j{\int}%
\global\long\def\w{}%
\global\long\def\R{\Rightarrow}%
\global\long\def\*{\times}%
\global\long\def\={\equiv}%
\global\long\def\.{\cdot}%

\global\long\def\cA{\mathcal{A}}%
\global\long\def\cD{\mathcal{D}}%
\global\long\def\cF{\mathscr{\mathcal{F}}}%
\global\long\def\cH{\mathcal{H}}%
\global\long\def\cL{\mathcal{L}}%
\global\long\def\cJ{\mathcal{J}}%
\global\long\def\cO{\mathcal{O}}%
\global\long\def\cP{\mathcal{P}}%
\global\long\def\cQ{\mathcal{Q}}%
\global\long\def\cY{\mathcal{Y}}%

\global\long\def\sA{\mathscr{A}}%
\global\long\def\sD{\mathscr{D}}%
\global\long\def\sF{\mathscr{F}}%
\global\long\def\sH{\mathscr{H}}%
\global\long\def\sL{\mathscr{L}}%
\global\long\def\sJ{\mathscr{J}}%
\global\long\def\sO{\mathscr{O}}%
\global\long\def\sP{\mathscr{P}}%
\global\long\def\sQ{\mathscr{Q}}%
\global\long\def\sY{\mathscr{Y}}%

\global\long\def\na{\nabla}%
\global\long\def\cd{\cdots}%
\global\long\def\da{\dagger}%
\global\long\def\ot{\otimes}%
\global\long\def\we{\wedge}%
\global\long\def\qu{\quad}%

\global\long\def\fL{\mathfrak{L}}%
\global\long\def\md{\mathrm{d}}%
\global\long\def\re{\mathrm{Re}}%
\global\long\def\im{\mathrm{Im}}%
\global\long\def\hb{\hbar}%

\title{Collapsing dust thin shells in Einstein-Gauss-Bonnet gravity
}


\author{Yong-Ming Huang\thanksref{e1,addr1}
	\and
	Yu Tian\thanksref{e2,addr1,addr2} 
	\and
	Xiao-Ning Wu\thanksref{e3,addr3,addr4,addr5}
}

\thankstext{e1}{e-mail: huangyongming15@mails.ucas.ac.cn}
\thankstext{e2}{e-mail: ytian@ucas.ac.cn}
\thankstext{e3}{e-mail: wuxn@amss.ac.cn}
\institute{School of Physical Sciences, University of Chinese Academy of Sciences, Beijing 100049, China \label{addr1}\and 
	Institute of Theoretical Physics, Chinese Academy of Sciences, Beijing 100190, China \label {addr2}\and
	Institute of Mathematics, Academy of Mathematics and System Science, Chinese Academy of Sciences, Beijing 100190, China \label{addr3} \and Hua Loo-Keng Key Laboratory of Mathematics, Chinese Academy of Sciences, Beijing 100190, China\label{addr4} \and School of Mathematical Sciences, University of Chinese Academy of Sciences, Beijing 100049, China \label{addr5}
}

\date{Received: date / Accepted: date}

\maketitle

\begin{abstract}
We investigate gravitational collapse of a spherically symmetric thin shell in the Einstein-Gauss-Bonnet (EGB) gravity. Under the recently proposed 4D limit, we find that the collapsing
shell will be bounced back at a small radius, without forming a singularity.
This bouncing behavior is similar to those of a test particle and
a homogeneous spherical dust star, in accordance with the expectation
that the Gauss-Bonnet term will modify the small scale behavior of
the Einstein gravity. We analyze the causal structure of the dynamic
spacetime that represents the bouncing process, finding that the thin
shell has an oscillation behavior on the Penrose diagram, which means
that the thin shell results in a novel type of black hole with respect
to observers outside the event horizon that the collapse forms. We also find that the weak cosmic censorship conjecture holds in this model. Further
implications of such a regular gravitational collapse are discussed.\black
\end{abstract}

\section{Introduction}
\label{sec:intro}

The EGB theory is one of the most promising candidates for modified
gravity theory. In the past several decades, higher dimensional EGB gravity
has been extensively studied, and it has been showed that the Gauss-Bonnet term yields richer
phenomena than Einstein theory. On the other hand, the
Gauss-Bonnet invariant makes no contribute to four-dimensional field equations because it is currently a total derivative in gravitational action.

Recently, a novel proposal for four-dimensional EGB gravity has been published \cite{Glavan:2019inb}. By rescaling the GB coupling constant $\a\rightarrow\frac{\a}{D-4}$, and
take $D\rightarrow4$ limit at equation of motion (EOM) level, the Gauss-Bonnet term generates non-trivial dynamics.  Remarkably,
the simple strategy circumvent the Lovelock\textquoteright s theorem
and be free from Ostrogradsky instability. On the basis of this approach to gravitational physics, a new branch of static
spherically symmetric black hole solutions was obtained in \cite{Glavan:2019inb},
which was also given by the gravity theory with conformal anomaly \cite{Cai:2009ua}, though $\a$ has different meanings
in these contexts. As a result of that proposal, the solution has
regained widespread attention, and various properties of the 4D EGB
solutions were considered within a short period time. For example, gravitational
collapse was considered in \cite{Malafarina:2020pvl}, the generalization
of the original solution \cite{Glavan:2019inb} was studied in \cite{Wei:2020ght,Kumar:2020owy,Fernandes:2020rpa,Ghosh:2020vpc,Konoplya:2020qqh},
black hole thermaldynamics was investigated in \cite{Hegde:2020xlv,Wei:2020poh}, gravitational
lensing and shadow were considered in \cite{Kumar:2020sag,Kumar:2020owy,EslamPanah:2020hoj,Guo:2020zmf,Wei:2020ght},
quasinormal modes and stability were showed in \cite{Konoplya:2020bxa,Mishra:2020gce,Zhang:2020sjh},
the electromagnetic radiation properties of thin accretion disk around
black hole were explored in \cite{Liu:2020vkh}, while the Hawking radiation and black hole evaporation were showed in \cite{Zhang:2020qam,Wu:2021zyl} respectively. 

As research progresses, the validity of 4D EGB gravity is being questioned \cite{Ai_2020,Arrechea:2020gjw,Gurses:2020ofy,Mahapatra:2020rds,Gurses:2020rxb,Dadhich:2020ukj,Hennigar:2020lsl,Lu:2020iav}. For example, in this scheme, the field equations are actually ill defined in terms of general spacetime geometry \cite{Gurses:2020ofy,Gurses:2020rxb}, and the scheme also results in divergence of black hole entropy \cite{Lu:2020iav}. However, it is necessary to note that the scheme works well in certain geometries, such as those with spherical symmetry, and provides a well-defined action \cite{Fernandes:2020rpa} in those cases. As a result, it is reasonable to treat the 4D EGB gravity with certain symmetry as an effective theory, allowing us to investigate the effect of higher-order curvature correction in four dimensions and shedding light on the validity of the underlying theory. On the other hand, the model presented in this paper also can be regarded as an effective model of gravity theory with conformal anomaly \cite{Cai:2009ua} from which the spherically symmetric solution of 4D EGB gravity can be deduced. In other words, the model is worth studying even in the absence of a complete theory of 4D EGB gravity.

Gravitational collapse is one of the central issues in gravitational physics, as it is intimately connected to a number of significant issues, including the formation of black hole, cosmic censorship, and black hole thermal dynamical laws. As a simple model, the final fate of spherically symmetric gravitational collapse of a dust fluid has been extensively studied in four and higher dimensions \cite{Christodoulou:1984mz, Newman:1985gt, Banerjee:2002sy, Goswami:2002he} for Einstein's gravity, which demonstrates that the property of singularity depends on dimensionality and initial data, while the results also hold for Gauss-Bonnet gravity \cite{Maeda:2006pm, Jhingan:2010zz, Ghosh:2010jm, Zhou:2011vy}, therefore it is worthwhile to consider the gravitational collapse in 4D EGB gravity. The spherical thin shell is an excellent toy model for studying the gravitational collapse process, which has been considered by several authors \cite{1979CMaPh..68..291F,Israel:1986gqz,Wald:1997wa,Hubeny:1998ga} and used to test many significant gravity-related conjectures, such as cosmic censorship conjecture.

In this paper, we investigate the motion of a spherically collapsing dust thin shell
in EGB gravity. It shows that when the gravitational mass $M$ and rest mass $m$ are within a certain parameter range, the collapsing thin shell on the Penrose diagram exhibits oscillating behavior in four dimensions. Then, using the method described in \cite{Boulware:1973tlq,Gao:2008jy}, we examine the thin shell's oscillating behavior on the Penrose diagram. Due to the fact that the thin shell would not collapse to form a black hole as a result of this behavior, it is critical to note that the bouncing behavior is not an effect observable to outside observers. The thin shell's motion in higher dimensions is also analyzed. Furthermore, the weak cosmic censorship conjecture is tested and turns out to hold in this model.

The paper is organized as follows: In Sec.~II, we introduce the bouncing behavior of collapsing dust stars in 4D EGB gravity. Then we are motivated to study collapsing
shells and obtain the EOM of spherical thin shells in $D$ dimensions in Sec.~III. In Sec.~IV, we study in detail the EOM of thin shells both in four
and higher dimensions. In Sec.~V, we test the weak cosmic censorship in this model. Finally, some concluding remarks will be presented in Sec.~VI.
\section{Dust collapse in the novel 4D EGB gravity}

In this section, we review the process of gravitational collapse in four-dimensional EGB gravity and show that collapsing dust stars exhibit  bouncing behavior. Actually, dust collapse in 4D EGB gravity has been studied
in \cite{Malafarina:2020pvl}, but they focused on the marginally bound
collapse of dust stars and did not consider the bouncing behavior. Because the surface of a dust star follows  geodesics of test particles with respect to the external spacetime (4D EGB black hole) \cite{Malafarina:2020pvl}, we concentrate on the geodesics rather than directly analyzing the collapsing dust star's trajectory . 

We first introduce the 4D EGB gravity. Consider $D\ge5$ dimensional EGB theory which has action
\[
S_{M}=\frac{1}{16\pi G_{d}}\int d^{D}x\sqrt{-g}\left(R+\a L_{GB}\right)
,\]where the Guass-Bonnet term $L_{GB}$ is defined by
\[
L_{GB}=R^{2}-4R_{ab}R^{ab}+R^{abcd}R_{abcd}.
\]

For this theory, the spherically symmetric vacuum solution has given by Boulware and Deser \cite{Boulware:1985wk},
\begin{equation}
	ds^{2}=-H(r)dt^{2}+\frac{dr^{2}}{H(r)}+r^{2}d\W_{D-2}^{2},\label{eq:1}
\end{equation}
where $d\W_{D-2}^{2}$ is the line element of the unit $S^{D-2}$ and
\[
H(r)=1+\frac{r^{2}}{2\widetilde{\a}}\left(1-\sqrt{1+\frac{16\widetilde{\a}M}{(D-2)r^{D-1}}}\right),
\]
where $\widetilde{\a}=(D-3)(D-4)\a$.

The recent proposal of 4D EGB gravity \cite{Glavan:2019inb} extends (\ref{eq:1}) to $D=4$ and gives four dimensional EGB vacuum solution, namely
\begin{equation}
	ds^{2}=-F(r)dt^{2}+\frac{dr^{2}}{F(r)}+r^{2}\left(d\theta^{2}+\sin^{2}\theta   d\phi^{2}\right),\label{eq:1-1}
\end{equation} where
\[F(r)=1+\frac{r^{2}}{2\a}\left(1-\sqrt{1+\frac{8\a M}{r^{3}}}\right),\]
then $F(r)=0$ gives the event horizons of 4D EGB black hole, namely
\begin{equation}
	r_{\pm}=M\pm\sqrt{M^2-\a}.\label{eq:1-2}
\end{equation}  
Notice that (\ref{eq:1-1}) is well-defined when $r\rightarrow0$,  whereas a straight-forward calculation demonstrates that the scalar $R^{abcd}R_{abcd}$ diverge as $\frac{M\a}{r^{3}}$ when $r\rightarrow 0$, implying that $r=0$ is a real singularity.

Following that, we will reveal the existence of bouncing behavior for a test particle which is freely falling in 4D EGB gravity. In this section, we restrict ourselves to non-extremal cases, where both $r_{+}$ and $r_{-}$ in (\ref{eq:1-2}) exist and have different values.

The radial geodesic equation is given by
\begin{equation}
	g_{rr}\left(\frac{dr}{d\t}\right)^{2}+g_{tt}\left(\frac{dt}{d\t}\right)^{2}=-1.\label{eq:2}
\end{equation}
 Considering the conserved quantities
associated with the Killing vectors $\partial_{t}$, there is
\begin{equation}
	E=-p_{t}=-g_{tt}p^{t}=-g_{tt}\frac{dt}{d\t}.\label{eq:3-5}
\end{equation}

Combine $(\ref{eq:2})$ with $(\ref{eq:3-5})$, we obtain
\begin{equation}
	\left(\frac{dr}{d\t}\right)^{2}=E^{2}-F(r),\label{eq:4-2}
\end{equation}
which tells us that the physical region of trajectory of test particles
should be confined in $E^{2}-F(r)\geq0$. Due to the fact that $F(r\rightarrow0)=1$, test particles with $E\in[0,1)$ must be bounced back before reaching $r=0$ .

By analyzing geodesics of test particles in a 4D EGB black hole background, we conclude that there exist bouncing behavior for  collapsing
dust stars. We then ask, is the bouncing behavior a universal phenomena
for self-gravitational collapsing system in 4D EGB theory? To explore this issue further, we are motivated to investigate the collapsing thin spherical shell in this theory.

\section{The equation of motion of thin shell in D dimensions}

In this paper, the model is based on a $D$ dimensional spherically symmetric
spacetime $M$ that is split into two segments by a timelike hypersurface $\S$ with two sides denoted $\S_{\pm}$. The hypersurfcae can then be treated as the boundary of each half of the spacetime. By varying the action, the generalized Israel junction condition \cite{Davis:2002gn} on $\S$ is obtained, which has been rigorously re-derived recently by expressing the field equations in terms of distributions \cite{Chu:2021uec}. Note that the hypersurface describes the path of a thin spherical shell. Our goal in this section is to derive the shell's EOM under the equation of state and junction condition. To begin, we consider D dimensional EGB theory with a spherically symmetric thin shell, whose action is given by
\begin{eqnarray}
	S=&&S_{M}+S_{matter}=\frac{1}{16\pi G_{D}}\int_{M}d^{D}x\sqrt{-g}\left(R+\a L_{GB}\right)\nonumber\\
	&&-\int_{\S}d^{D-1}x\sqrt{-h}L_{m}^{\S},\label{eq:5-2}
\end{eqnarray} where $h_{ab}$ is the induced metric on $\S$.

As in the Einstein case, a surface term must be added to \eqref{eq:5-2} to obtain a well-defined variational problem. The corresponding term is
\begin{eqnarray}
	S_{\S}=-\frac{1}{8\pi G_{D}}\int_{\S_{\pm}}d^{D-1}x\sqrt{-h}\left(K+2\a\left(J-2\hat{G}^{ab}K_{ab}\right)\right), \nonumber 
\end{eqnarray}
where $\hat{G}^{ab}$ is the Einstein tensor related to $h_{ab}$,
$J$ is the trace of $J_{ab}$ which defined as
\begin{eqnarray}
	J_{ab}=&&\frac{1}{3}(2KK_{ac}K_{b}^{c}+K_{cd}K^{cd}K_{ab}-2K_{ac}K^{cd}K_{db}-K^{2}K_{ab}), \nonumber 
\end{eqnarray}
where $K$ is the trace of extrinsic curvature, and the extrinsic curvature
of thin shell is given by \footnote{$\m,\n$ are the indexes of coordinates of spacetime, while $a,b$
	are the indexes of coordinates of $\S$.}
\begin{equation}
	K_{ab}=h_{a}^{\m}h_{b}^{\n}\nabla_{\m}n_{\n},\label{eq:6-2}
\end{equation} where $n^{\n}$ is the normal vector of the thin shell.

As an aid to derive junction conditions, we introduce Gaussian
normal coordinates in the neighborhood of $\S$. The metric
$g_{\m\n}$ has the following form
\begin{eqnarray}
	ds^{2}=&&d\o^{2}+h_{ab}dx^{a}dx^{b}=d\o^{2}-n^{2}(\o,\t)d\t^{2}  \nonumber \\ 
	&&+\stackrel[i=1]{D-2}{\sum}\frac{r^{2}\left(\o,\t\right)}{\left(1+\frac{1}{4}\stackrel[j=1]{D-2}{\sum}x_{j}^{2}\right)^{2}}dx_{i}^{2}.\label{eq:7-2}
\end{eqnarray} Note that the extrinsic curvature of surfaces $\o=$ constant is $K_{ab}=-\frac{1}{2}\6_{\o}h_{ab}$.
Substituting the metric ansatz (\ref{eq:7-2}) into $S_{tot}=S+S_{\S}$,
then $S_{tot}$ reduces to the following form
\begin{eqnarray}
	S_{reduced}=&&\frac{A_{D-2}}{16\pi G_{D}}\int d\o d\t (D-2)nr^{D-3}\Bigg(-2\6_{\o}^{2}r  \nonumber\\
	&&+(-3+D)\frac{\psi}{r}+\frac{4\widetilde{\a}}{r^{2}}\bigg(-\psi\6_{\o}^{2}r+\frac{1}{2r}(-5+D) \nonumber\\ 
	&&\Big(\frac{1}{2}-\left(\6_{\o}r\right)^{2}-\frac{\dot{r}^{2}}{n^{2}}\left(\psi+\frac{7\dot{r}^{2}}{6n^{2}}\right) +\frac{1}{2}\left(\6_{\o}r\right)^{4}\Big)\bigg)\Bigg) \nonumber \\
	 &&+\frac{A_{D-2}}{8\pi G_{D}}\int_{\S_{\pm}}d\t(D-2)nr^{D-3}\6_{\o}r\Bigg(1+\frac{2\widetilde{\a}}{r^{2}} \nonumber \\
	&&\left(\psi+\text{\ensuremath{\frac{2}{3}\left(\6_{\o}r\right)^{2}}}\right)\Bigg)-\int_{\S}d^{D-1}x\sqrt{-h}L_{m}^{\S},  \label{eq:8-2} 
\end{eqnarray} 
where $A_{D-2}=\frac{2\pi^{\frac{D-1}{2}}}{\G[\frac{D-1}{2}]}$ is
the area of unit $S^{D-2}$, $\dot{r}$ denotes $\6_{\t}r$ and 
\[
\psi=1-\frac{\dot{r}^{2}}{n^{2}}-\left(\6_{\o}r\right)^{2}.
\]

Varying the reduced action (\ref{eq:8-2}), one can obtain junction
condition
\begin{eqnarray}
	&&\frac{(D-2)}{n^{2}}\left[K_{i}^{i}\left(1+2\widetilde{\a}\left(\frac{1}{r^{2}}+\frac{\dot{r}^{2}}{n^{2}r^{2}}-\frac{1}{3}\left(K_{i}^{i}\right)^{2}\right)\right)\right]_{-}^{+}\nonumber\\
	&&=-8\pi G_{D}S^{\t\t}<+\infty,\label{eq:9}
\end{eqnarray} where $\left[X\right]_{-}^{+}\doteq X_{+}-X_{-}$ (we have chosen the normal vector of
$\S$ to be outward-pointing, i.e pointing from the inside of shell
to outside) and the energy-momentum tensor
is defined by 
\[S^{ab}=\frac{2}{\sqrt{-h}}\frac{\d S_{matter}}{\d h_{ab}}.\]
Note that $K^i_{i}=\frac{1}{D-2}H^{ij}K_{ij}$, where $H_{ij}dx^{i}dx^{j}=h_{ab}dx^{a}dx^{b}+n^{2}(\omega,\t)d\t^2.$

Suppose the velocity of an comoving observer on the radial collapsing thin shell to be $u^{a}=(\6_{\t})^a$, where $\t$ is the proper time of the observer,
then the metric of the shell has the form 
\begin{equation}
	ds_{D-1}^{2}=-d\t^{2}+r^{2}(\t)d\W_{D-2}^{2}.\label{eq:10-1}
\end{equation}

Assuming that the shell satisfies pressureless condition, its surface energy momentum has 
\begin{equation}
	S^{ab}=\text{\ensuremath{\s}(\ensuremath{\t})}u^{a}u^{b},\label{eq:9-3}
\end{equation}
where $\s$ denotes surface density. From the conservation
equations  $^{(D-1)}\nabla_{b}S_{a}^{b}=0$, where
$^{(D-1)}\nabla_{b}$ is the derivative operator on $\text{\ensuremath{\S}}$,
we obtain
\begin{equation}
	\frac{\dot{\s}}{\s}+(D-2)\frac{\dot{r}}{r}=0.\label{eq:11}
\end{equation} The rest mass of shell is defined as $m=\s{A_{D-2}}r^{D-2}$ and ($\ref{eq:11}$) would implies $m$ is a constant.

Since we consider a thin shell, vacuum condition holds in inner and outside of the shell, therefore the metric of bulk is given by
\begin{equation}
	ds^{2}=-f(r)dt^{2}+\frac{dr^{2}}{f(r)}+r^{2}d\W_{D-2}^{2},\label{eq:10-2}
\end{equation} where 
\begin{eqnarray}
	f(r)&\equiv& f_{-}(r)=1,\nonumber \\
	f(r)&\equiv& f_{+}(r)=1+\frac{r^{2}}{2\widetilde{\text{\ensuremath{\a}}}}\left(1-\sqrt{1+\frac{16\widetilde{\a}M}{(D-2)r^{D-1}}}\right) \nonumber.
\end{eqnarray}
Because both sides must have the same induced metric on the shell, then we have 
\[
ds_{\S}^{2}=\left(-f_{\pm}(r)\dot{t}^{2}+\frac{1}{f_{\pm}(r)}\dot{r}^{2}\right)d\t^{2}+r^{2}(\t)d\W_{D-2}^{2},
\] where
\begin{equation}
	-f_{\pm}(r)\dot{t}^{2}+\frac{1}{f_{\pm}(r)}\dot{r}^{2}=-1.\label{eq:3}
\end{equation} The equation ($\ref{eq:3}$) means if we know the $r(\t)$, then we would
obtain $t(\t)$ so that the spherical shell's motion $r(t)$ is clear for observer at infinity.

Now we define the hypersurface $\S$ as $r=r(\t)$, which describes the motion of thin shell, then its tangent vector $u^a$ can be written in bulk coordinates as
\[
u^a=\dot{t}(\6_{t})^a+\dot{r}(\6_{r})^a 
\] and its normal vector 
$n^{a}$ has
\[
n^{a}=n^{t}\left(\frac{\text{\ensuremath{\6}}}{\6 t}\right)^{a}+n^{r}\left(\frac{\text{\ensuremath{\6}}}{\6 r}\right)^{a}.
\]

Combine $n^{a}n_{a}=1$ with $u^{a}n_{a}=0$, we can obtain
\begin{equation}
	n^{r}=\pm\sqrt{f(r)+\dot{r}^{2}},\label{eq:11-3}
\end{equation}
where $\pm$ determine the direction of $n^{a}$. More specifically,
we write $n^{r}$ outside and inside the shell as
\begin{eqnarray}
	n_{o}^{r}&=&\pm\sqrt{f_{+}(r)+\dot{r}^{2}},\nonumber \\
	n_{i}^{r}&=&\sqrt{1+\dot{r}^{2}}. \nonumber
\end{eqnarray}

Due to the fact that the interior spacetime is flat, $n_{i}^a$ points to increasing $r$ in our convention, which implies that $n_{i}^{r}$ should be positive.

Then, from $(\ref{eq:6-2})$,(\ref{eq:10-1}),(\ref{eq:10-2}), one can obtain 
\begin{equation}
	K_{i}^{i}=r^{-1}n^{r},\label{eq:11-4}
\end{equation} and, when combined with $(\ref{eq:9}),(\ref{eq:10-1}),(\ref{eq:9-3})$, we
conclude the EOM of shell
\begin{eqnarray}
	&&\frac{(D-2)}{8\pi G_{D}r}\big(\left(n_{i}^{r}-n_{o}^{r}\right)+\frac{2\widetilde{\a}}{3r^{2}}\big(3(1+\dot{r}^{2})(n_{i}^{r}-n_{o}^{r})\nonumber \\&&
	+\left(n_{o}^{r}\right)^{3}-\left(n_{i}^{r}\right)^{3}\big)\big) =S_{\t\t}=\s.\label{eq:3-1}
\end{eqnarray} 

For simplicity, we choose the following units for the rest mass of shell
\[
\frac{8\pi G_{D}}{(D-2)A_{D-2}}=1,
\] then equation (\ref{eq:3-1}) becomes
\begin{eqnarray}
	&&r^{D-3}\big(\left(n_{i}^{r}-n_{o}^{r}\right)+\frac{2\widetilde{\a}}{3r^{2}}\big(3(1+\dot{r}^{2})(n_{i}^{r}-n_{o}^{r})\nonumber \\
	&&+(n_{o}^{r})^{3}-(n_{i}^{r})^{3}\big)\big)=m,\label{eq:4-4}
\end{eqnarray}
while we have $(n_{i}^{r})^2=1+\dot{r}^{2}$ in our case, therefore (\ref{eq:4-4}) can be written as
\begin{eqnarray}
	&&r^{D-3}\big((n_{i}^{r}-n_{o}^{r})+\frac{2\widetilde{\a}}{3r^{2}}\big(n_{o}^{r}\left(\left(n_{o}^{r}\right)^{2}-3\left(n_{i}^{r}\right)^{2}\right)\nonumber \\
	&&+2\left(n_{i}^{r}\right)^{3}\big)\big)=m.\label{eq:4}
\end{eqnarray} 
Notice that $\widetilde{\a}$ has units $[L^{2}]$.
We replace $r\rightarrow\widetilde{\a}^{\frac{1}{2}}r,m\rightarrow\widetilde{\a}^{\frac{D-3}{2}}m,M\rightarrow\widetilde{\a}^{\frac{D-3}{2}}M$ and have dimensionless equation
\begin{eqnarray}
	&&r^{D-3}\big((n_{i}^{r}-n_{o}^{r})+\frac{2}{3r^{2}}\big(n_{o}^{r}\left(\left(n_{o}^{r}\right)^{2}-3\left(n_{i}^{r}\right)^{2}\right)\nonumber \\
	&&+2\left(n_{i}^{r}\right)^{3}\big)\big)=m,\label{eq:16}
\end{eqnarray}
while $f_{+}(r)=1+\frac{r^{2}}{2}(1-\sqrt{1+\frac{16M}{(D-2)r^{D-1}}})$.  In following sections, we always take EOM and $f_{+}(r)$ the form which regard $\{r,m,M\}$ as dimensionless parameters when we refer to EOM and $f_{+}(r)$ in specific dimensions.

\section{Analysis of collapsing thin shells' EOM}

In this section, we will analyze the EOM of spherical shells. The existence of bounce behavior at small radius is firstly shown for our spherically symmetric model based on 4D EGB gravity, and then the trajectories with bounce behavior are classified into three types based on their difference on the Penrose diagram. Finally, we will examine the trajectory of thin shells and show that the bouncing behavior is absent in higher dimensions. For $D=4$, we make rescaling $\a\rightarrow\frac{1}{D-4}\a$, which corresponds to $\widetilde{\a}=(D-3)\a$. 

\subsection{The bouncing process in $D=4$}

Let us begin by demonstrating the central point of this paper, namely the existence of bouncing behavior for a dust thin shell. In this subsection, we restrict to non-extremal cases. Equation (\ref{eq:16}) can be re-written as
\begin{eqnarray}
	&&\dot{r}^{6}+\left(\frac{3}{2}r^{2}+3-\frac{M^{2}}{m^{2}}\right)\dot{r}^{4}+\left(\frac{3}{4}r^{2}+1\right)^{2}-\frac{9W^{2}}{64m^{2}}\nonumber \\&&
	+\left(\left(\frac{3}{4}r^{2}+2\right)^{2}-1-\frac{3MW}{4m^{2}}\right)\dot{r}^{2}=0,\label{eq:19}
\end{eqnarray}
where
\begin{eqnarray}
	W&=&m^{2}r+\frac{1}{18}\left(-1+\sqrt{1+\frac{8M}{r^{3}}}\right)r^{5} \nonumber\\
	&&+\frac{4}{9}M\left(6+\left(3+\sqrt{1+\frac{8M}{r^{3}}}\right)r^{2}\right). \nonumber \end{eqnarray}.

Eq.(\ref{eq:19}) is a cubic equation of $\dot{r}^{2}$. According to the discriminant of the cubic algebraic
equation, there is a unique real solution to (\ref{eq:19}), which can be written as 
\begin{eqnarray}
	&&\dot{r}^{2}=V(r,m,M),\\
	&&V(r,m,M)=\left(-\frac{q}{2}+\sqrt{\left(\frac{q}{2}\right)^{2}+\left(\frac{p}{3}\right)^{3}}\right)^{\frac{1}{3}} \nonumber\\
	&&+\left(-\frac{q}{2}-\sqrt{\left(\frac{q}{2}\right)^{2}+\left(\frac{p}{3}\right)^{3}}\right)^{\frac{1}{3}}+\frac{M^{2}}{3m^{2}}-\frac{1}{2}r^{2}-1,\nonumber
\end{eqnarray}
where 
\begin{eqnarray}
	p&=&-\frac{M^{4}}{3m^{4}}+\frac{(2+r^{2})M^{2}}{m^{2}}-\frac{3WM}{4m^{2}}-\frac{3}{16}r^{4},\nonumber\\
	q&=&-\frac{9W^{2}}{64m^{2}}+\frac{\left(-2M^{3}+3m^{2}M(2+r^{2})\right)W}{8m^{4}}-\frac{2M^{6}}{27m^{6}}\nonumber\\
	&+&\frac{(2+r^{2})M^{4}}{3m^{4}}-\text{\ensuremath{\frac{(16+16r^{2}+5r^{4})M^{2}}{16m^{2}}-\frac{r^{6}}{32}}} \nonumber
\end{eqnarray}
and $V(r,m,M)\geq0$
corresponds to the physical allowable region of the spherical shell
trajectory. For illustrating the existence of the bouncing process,
without loss of generality, we numerically solve $V(r)=0$ for $M=5$ with different $m$ to see the number of turning points in the trajectory of spherical shell, see Figure 1. 

\begin{figure}
	\includegraphics[scale=0.66]{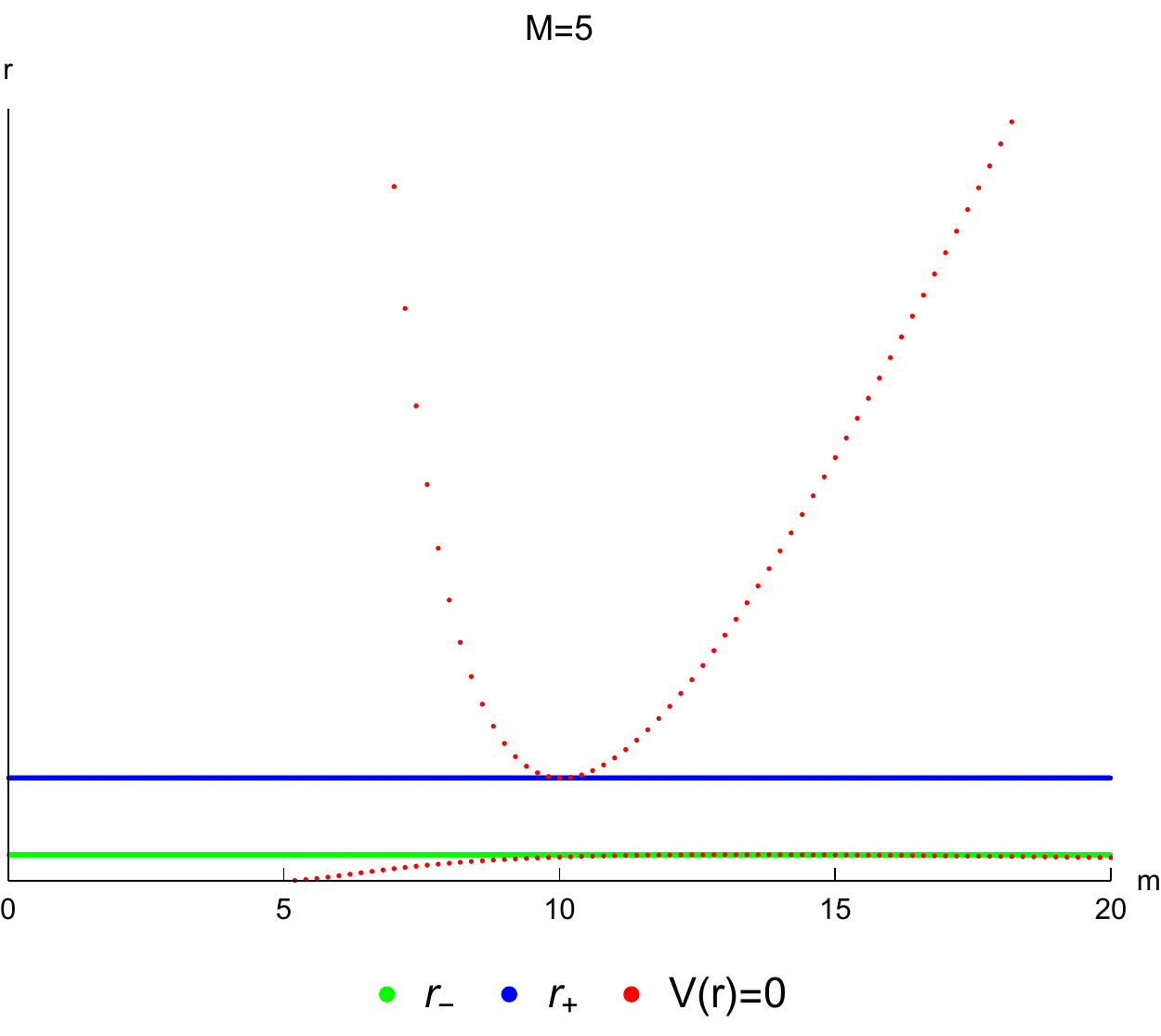}\caption{For $M=5$, the solutions of the equation $V(r)=0$ are plotted over the range $m=5$ to $m=20$, with the intervals $\protect\D m=0.2$. Note that the distance between horizons has been scale down to make it easier to display turning points.}
\end{figure}

To determine the trajectory of bouncing process more specifically, we would follow the analysis method described in \cite{Boulware:1973tlq,Gao:2008jy}.  Due to the fact that $\sqrt{f_{+}+\dot{r}^{2}}\geq0$ in equation (\ref{eq:16}) requires turning points to be located within the interval  satisfying $f_{+}\geq0$, since $f_{+}<0$
in $r_{-}<r<r_{+}$, there are no turning points for thin shells in this
interval. Supposing the trajectory of a spherical shell has two turning points which locate at $0<r_{1}<r_{-}$ and $r_{2}>r_{+}$ respectively, we will always refer to an oscillating trajectory as such in this paper. We now classify such trajectories into four types on the Penrose diagram. The construction of the Penrose diagram for 4D EGB black hole (\ref{eq:1-1}) is given in the Appendix, where also contains an illustration of the sign of $n_{o}^{r}$ on the diagram. To explain the classification rules, suppose a thin shell that starts in region $I_{+}$ in Figure 5, then the worldline of the shell will pass region $II_{+}$ to reach the minimum $r_{1}<r_{-}$. However, there are two possible ways to reach the minimum: by entering region $III_{+}$ or by entering region $III_{-}$. We then define the following two distinct types of trajectories: type $I$, which oscillates between regions $I_{+}$ and $III_{+}$, and type $II$, which oscillates between regions $I_{+}$ and $III_{-}$. Similarly, for a shell that begins in region $I_{-}$, its worldline will pass through region $II_{+}$ and then enter region $III_{+}$ or $III_{-}$ in order to reach minimum $r_{1}$, resulting the other two types of trajectories: type $III$, which oscillates between regions $I_{-}$ and $III_{+}$, and type $IV$, which oscillates between regions $I_{-}$ and $III_{-}$. For instance, the trajectories of type $I$ and type $IV$ on the Penrose diagram are shown in Figure 2. The four possible oscillation types and the sign of $n_{o}^{r}$ in each corresponding region are listed in Table 1.

\begin{figure*}
	\centering
	\subfloat[type $I$ trajectory]{\includegraphics[scale=0.65]{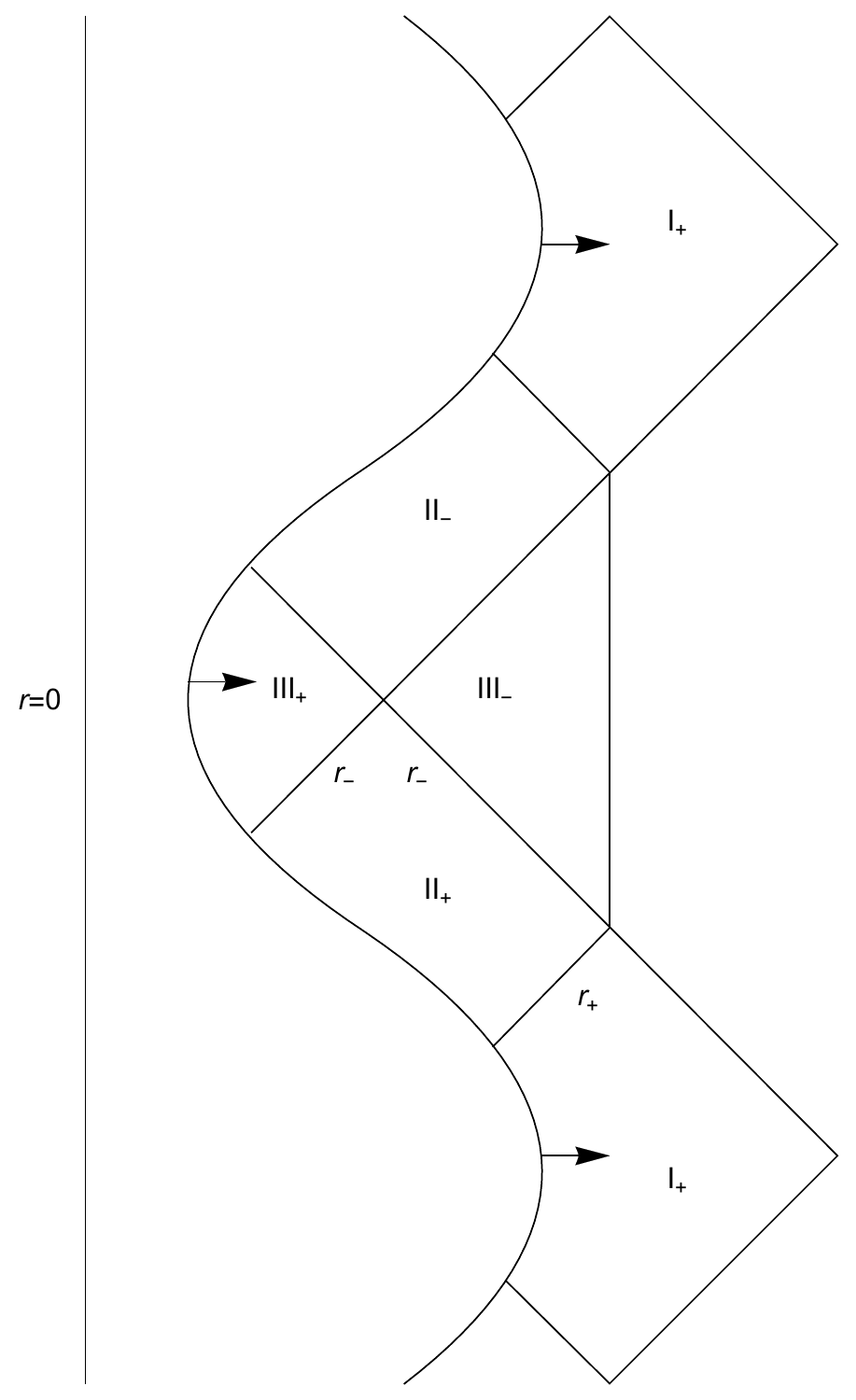}}\
	\subfloat[type $IV$ trajectory]{\includegraphics[scale=0.65]{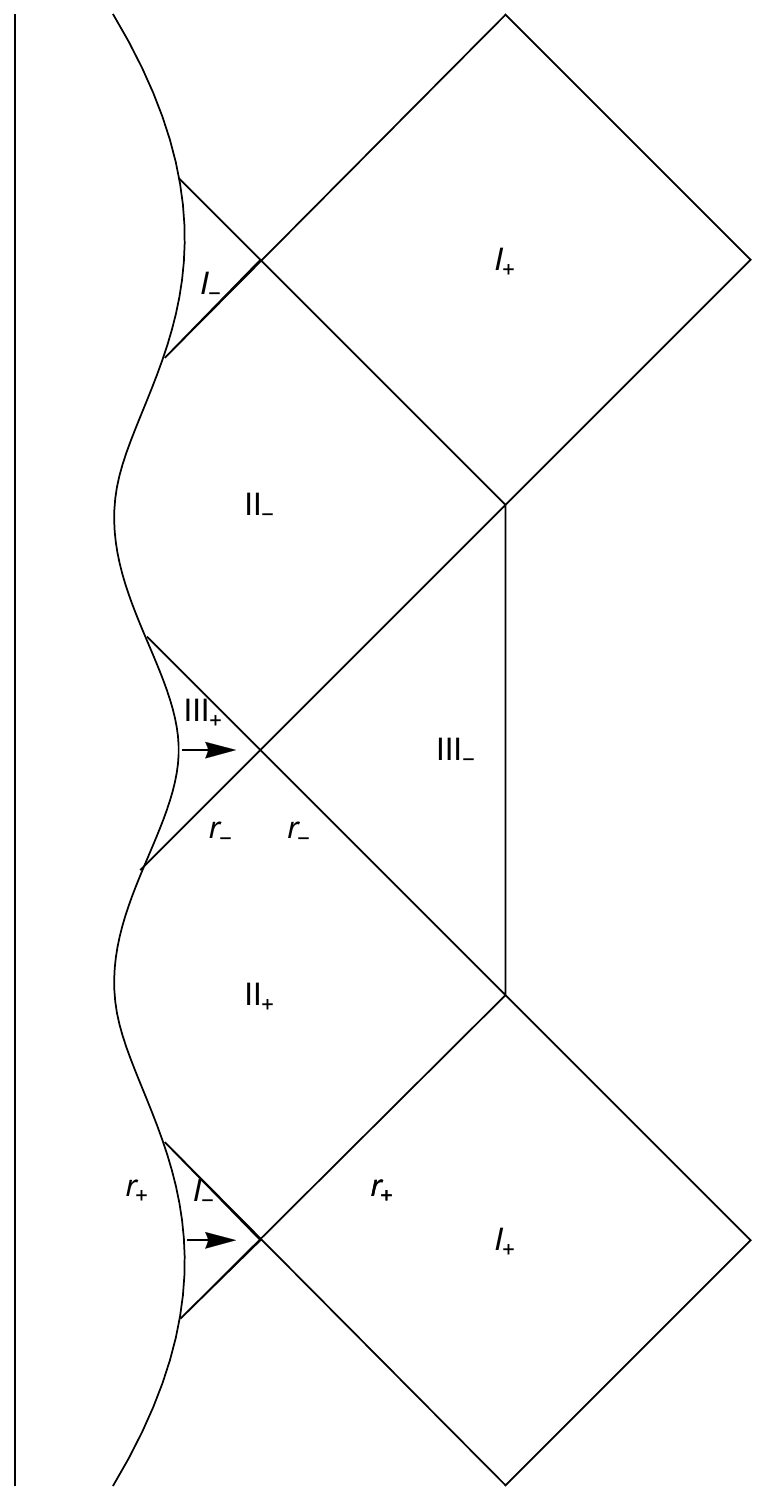}}\caption{The different types of the oscillating shell's trajectory on Penrose diagram are plotted.}
\end{figure*}

\begin{table}[htbp]
	\renewcommand\arraystretch{1.7}
	\setlength{\tabcolsep}{7mm}{
		\begin{tabular}{ccc}
			\hline\hline
			\textbf{Types} &$0<r\leq r_{-}$&  $r\geq r_{+}$   \\
			\hline
			\textbf{$I$} & $III_{+};n_{o}^{r}>0$& $I_{+};n_{o}^{r}>0$  \\
			\hline
			\textbf{$II$} &$III_{-};n_{o}^{r}<0$ & $I_{+};n_{o}^{r}>0$ \\
			\hline
			\textbf{$III$} & $III_{+};n_{o}^{r}>0$ & $I_{-};n_{o}^{r}<0$    \\
			\hline
			\textbf{$IV$} &$III_{-};n_{o}^{r}<0$ & $I_{-};n_{o}^{r}<0$   \\
			\hline
			\hline
	\end{tabular}}
	\caption{The four possible types of oscillation trajectories on the Penrose diagram, as well as each corresponding region, are listed.}
	\label{tab:dof_classes}
\end{table}

Now we ask such a question: which type of trajectory will an oscillating shell with specific mass parameters follow? To begin answering the question, we use type $I$ trajectory as an example to illustrate our method. Notice that the position of the turning points of the trajectory is decided by (\ref{eq:16}) with  $\dot{r}=0$. Specifically, the position of turning points with $n_{o}^{r}>0$ is determined by $g(r)=0$, where
\begin{equation}
	g(r)=r\left(1-\sqrt{f_{+}}\right)+\frac{2}{3r}\left(\sqrt{f_{+}}(-3+f_{+})+2\right)-m.\label{eq:6}
\end{equation}
Similarly, the position of turning points with $n_{o}^{r}<0$ is decided by $h(r)=0$, where
\begin{equation}
	h(r)=r\left(1+\sqrt{f_{+}}\right)+\frac{2}{3r}\left(-\sqrt{f_{+}}(-3+f_{+})+2\right)-m.\label{eq:22}
\end{equation}
Supposing the trajectory of an oscillating shell with mass parameters $\{m_{1}, M_{1}\}$ belongs to type $I$ trajectory whose two zero points with $n_{o}^{r}>0$. That means $g(r)=0$ with $\{m_{1}, M_{1}\}$ must have one zero point at $0<r<r_{-}$ and $r>r_{+}$ respectively, then analyzing the function $g(r)$ will reveal that the existence of such zero points distribution of $g(r)$ is equivalent to satisfying $1<M_{1}<m_{1}<M_{1}^{-}$, where $M_{1}^{-}=\frac{7}{3}M_{1}-\frac{1}{3}\sqrt{M_{1}^{2}-1}.$ Finally, we conclude that the trajectory of an oscillating shell with $1<M_{1}<m_{1}<M_{1}^{-}$ is classified into type $I$ trajectory. Similar analysis can be done for other type trajectories. The following is a detailed analysis:

The derived function of $g(r)$ and $h(r)$ are given by
\begin{eqnarray}
	g'(r)&=&1-\sqrt{f_{+}}+\frac{1}{r^{2}}\Big(f_{+}^{-\frac{1}{2}}u\left(f_{+}-1-\frac{3Mr}{u}\right)\nonumber\\
	&&-\frac{2}{3}\left(\sqrt{f_{+}}\left(f_{+}-3\right)+2\right)\Big),\nonumber\\
	h'(r)&=&1+\sqrt{f_{+}}-\frac{1}{r^{2}}\Big(f_{+}^{-\frac{1}{2}}u\left(f_{+}-1-\frac{3Mr}{u}\right)\nonumber\\
	&&-\frac{2}{3}\left(\sqrt{f_{+}}\left(f_{+}-3\right)-2\right)\Big), \nonumber
\end{eqnarray}
where $u=2(f_{+}-1)-r^{2}$. Then, by examining the sign of the derived functions, one can determine the monotonicity of $g(r)$ and $h(r)$ at various radial radius values, as illustrated in Table 2.
\begin{table}
	\renewcommand\arraystretch{1.7}
	\setlength{\tabcolsep}{7mm}\begin{tabular}{ccc}
		\hline \hline
		\textbf{Function} & $0<r\leq r_{-}$ & $r\geq r_{+}$\tabularnewline
		\hline 
		$g'(r)$ & $>0$ & $<0$\tabularnewline
		\hline 
		$h'(r)$ & $<0$ & $>0$\tabularnewline
		\hline \hline
	\end{tabular}\ %
	\caption{ The properties of the derivative functions $g(r)'$ and
		$h(r)'$ at $\{r|r>r_{+} \cup r_{-}>r>0\}$ are listed.}
\end{table} In order to analyze the existence of zero points of $g(r)$ and $h(r)$, we list boundary asymptotic behavior of these curves
\begin{eqnarray}
	&&g(r\rightarrow0)=-(m-M)+\frac{\sqrt{2}M^{\frac{3}{2}}r^{\frac{1}{2}}}{3}+O\left(r\right),\label{eq:7}\\
	&&g(r\rightarrow+\8)=-(m-M)+\frac{M^{2}}{2r}+O\left(r^{-2}\right),\label{eq:8}\\
	&&h(r\rightarrow0)=\frac{8}{3r}+(-m-M)+O\left(r^{\frac{1}{2}}\right),\label{25}\\
	&&h(r\rightarrow+\8)=2r+(-m-M)+O\left(r^{-1}\right).\label{eq:26}
\end{eqnarray}

(a). For $m>M$, the sign of $g(r)$ and $h(r)$ at their boundaries can be determined by their boundary asymptotic behavior. Combine with the monotonicity of the function $g(r)$ and $h(r)$ in $\{r|r>r_{+}\cup r_{-}>r>0\}$, the number of zero points for $g(r)$ and $h(r)$ in  $\{r|r>r_{+}\cup r_{-}>r>0\}$ depends on the sign of their values at horizons, there are four possibilities when there are two turning points for trajectory of thin shell and they show as
\begin{equation}
	\{M,m\}=A\cup B\cup C\cup D,\label{eq:27}
\end{equation}
where 
\begin{eqnarray}
	A&=&\{M,m|m>M \cap g(r_{\pm})>0\} \nonumber\\
	&=&\{M,m|M\geq1\cap M<m<M^{-}\}, \nonumber\\
	B&=&\{M,m|m>M \cap h(r_{\pm})<0\} \nonumber\\
	&=&\{M,m|M\geq1\cap m>M^{+}\}, \nonumber\\
	C&=&\{M,m|m>M \cap h(r_{-})<0\cap g(r_{+})>0\}=\emptyset, \nonumber\\
	D&=&\{M,m|m>M \cap h(r_{+})<0 \cap g(r_{-})>0\} \nonumber\\
	&=&\{M,m|M>1\cap M^{-}<m<M^{+}\}, \nonumber
\end{eqnarray}
where $M^{+}=\frac{7M}{3}+\frac{1}{3}\sqrt{M^{2}-1}.$ As a result, we conclude that the trajectory of an oscillating shell with mass parameters A/B/D is type I/IV/III, while type II trajectory is absent. To illustrate the classification in mass parameters'space in a more intuitive manner, we plot the Figure 3a. Particularly, at the
mass parameter $\{M=1,m=\frac{7}{3}\}$, we also find that the EOM
can describe a static configuration of the spherical shell, which is an extremal black hole for an outside observer. The shell locates exactly at the radial position of the horizon, as illustrated in Figure 3b.

\begin{figure*}
	\subfloat[]{\includegraphics[scale=0.65]{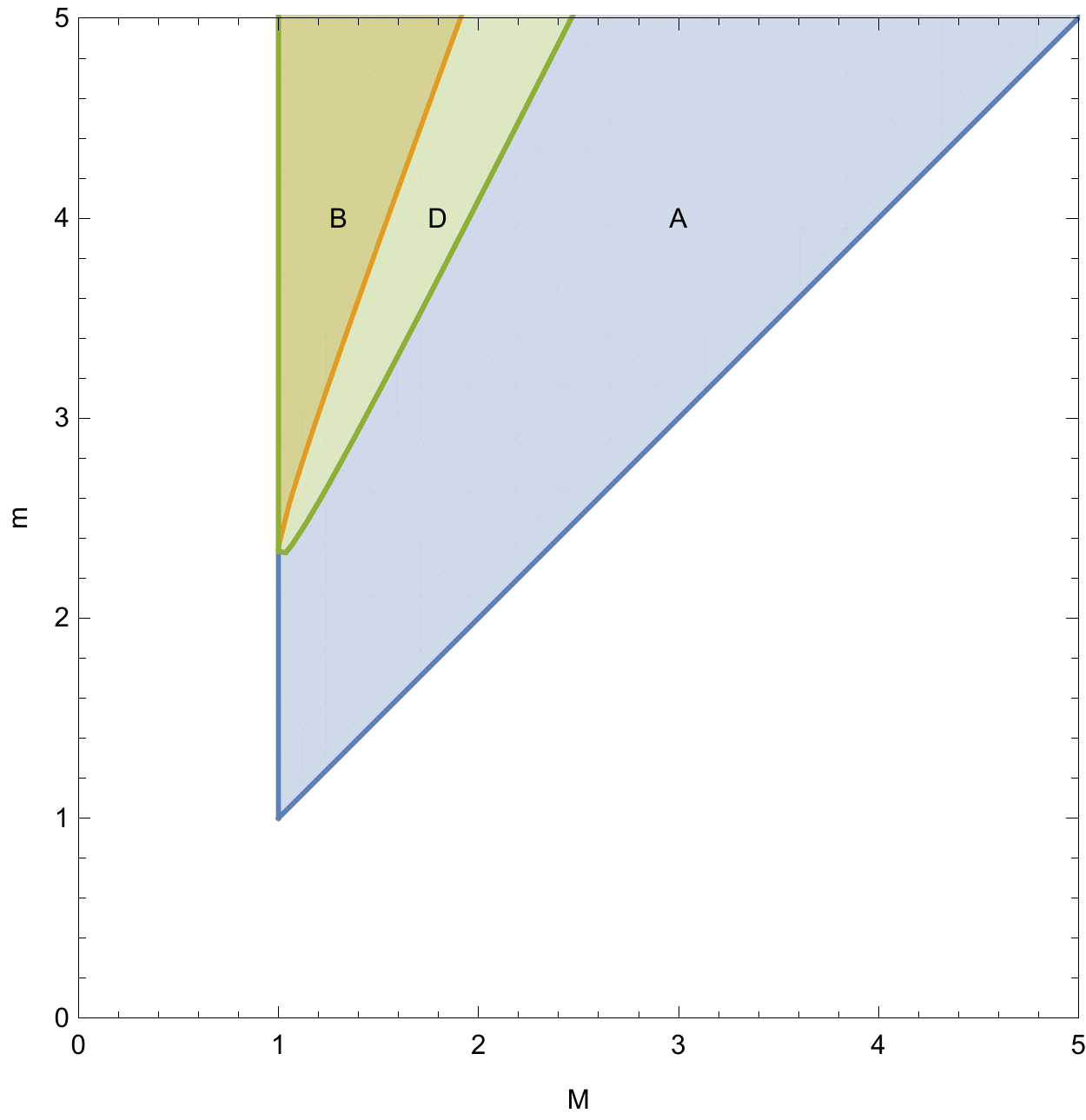}}\ \subfloat[]{\includegraphics[scale=0.71]{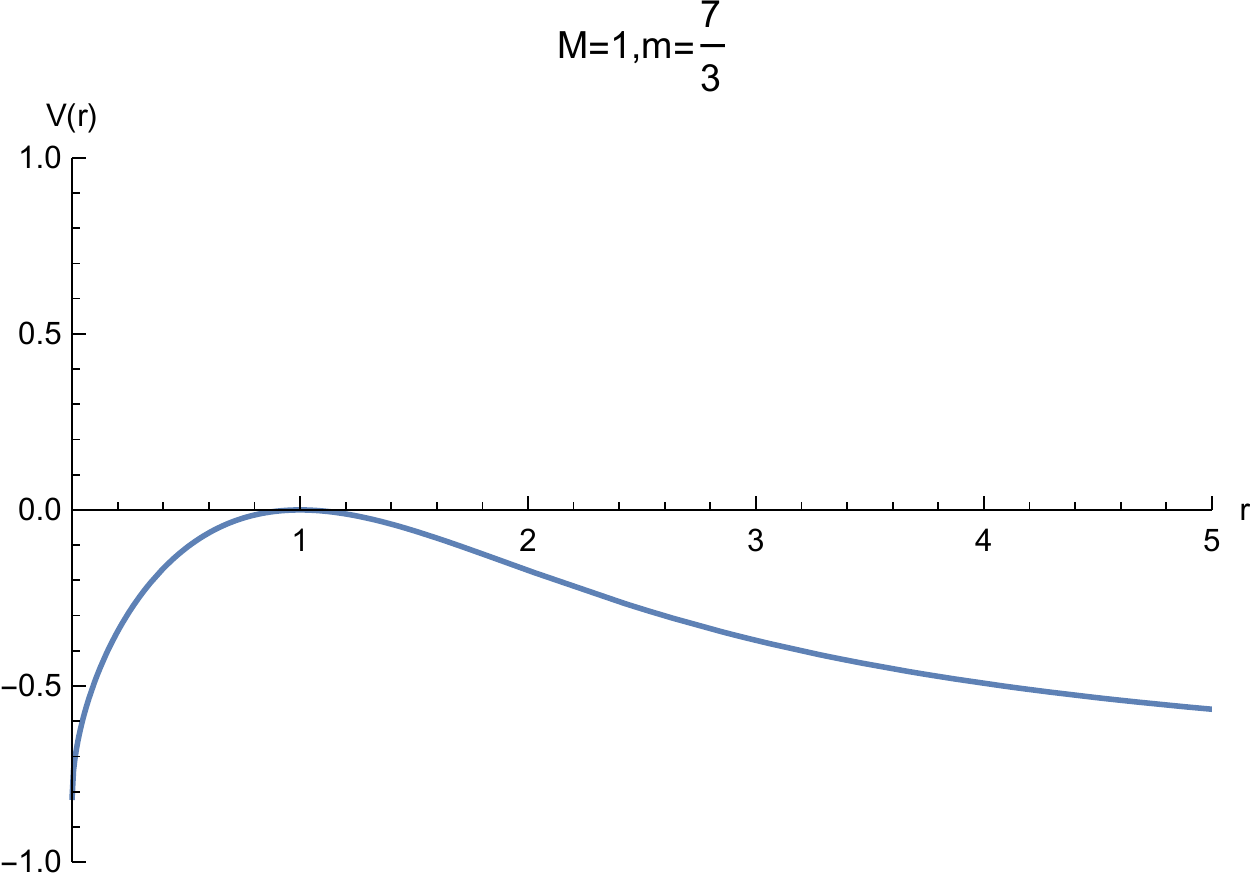}}\caption{(a). the classifications of bouncing processes by different mass parameters' regions.
		(b). the $V(r)$ profile of $\{M=1,m=\frac{7}{3}\},$ which corresponds to the intersection of region $A,B,D$ of left diagram. It shows that the spherical shell locate at $r=1$, where the position of horizon of extremal black hole is. } 
\end{figure*}

Until now, we have classified the oscillating thin shell
on the Penrose diagram into three cases, which means that we can infer the interval of the oscillating shell's mass parameter simply by observing its trajectory on the Penrose diagram. Physically, what we are actually interested in is the type I trajectory, whose mass parameters correspond to the region $A$ in Figure 3. This is because it is observable for the asymptotic observer in region $I_{+}$ on the Penrose diagram, which is not the case for type $III,IV$ trajectory, see Figure 2.  

(b). For $m\leq M$, we have $g(r\to0)>0$ and $g(r\to+\8)>0$. Due to the monotonic property and asymptotic behavior of $g(r)$ in $\{r|r>r_{+} \cup r_{-}>r>0\}$, one can deduce that $g(r)$ has no zero points, implying that the trajectory of a spherical shell does not have any turning points with $n_{o}^{r}>0$. Additionally, it also can be demonstrated that the trajectory of the spherical shell in this mass interval does not have any turning points with $n_{o}^{r}<0$. If we have such turning point,  it is equivalent to the mass parameters are either $\{M,m|h(r_{+})<0\}=\{M,m|M\geqslant1\cap m>\frac{7M}{3}-\frac{1}{3}\sqrt{M^{2}-1}\}$ or $\{M,m|h(r_{-})<0\}=\{M,m|M\geqslant1\cap m>\frac{7M}{3}+\frac{1}{3}\sqrt{M^{2}-1}\}$, which intersect with $m\leq M$ is the empty set, therefore we infer that the turning point of a shell's trajectory with $n_{o}^{r}<0$ is absent. Thus far, we have proved that the trajectory of a shell with $m\leq M$ does not have turning points, ie. the collapsing spherical shell with $m\leq M$ will eventually collapse into a singularity.
\textcolor{black}{{}}

\subsection{Collapsing shells in higher dimensions}

Now we turn our attention to the trajectory of a thin shell in higher dimensions. In this case, the position of horizon is determined by  $r^{D-5}\left(1+r^{2}\right)=\frac{4M}{D-2}$. Due to the fact that the left-hand side of this equation is monotonically increasing from zero to infinity for $D>5$, there is unique solution, implying the existence of single horizon. When $D=5$, the existence of horizon depends on the sign of $\frac{4}{3}M-1$, and there exist single horizon when $M>\frac{3}{4}$, which is the case here. 

Physically, we believe that the existence of bouncing behavior in four dimensions is due to its unique causal structure in comparison to that of higher dimensions. Indeed, because the Gauss-Bonnet term modifies the small scale behavior of the Einstein gravity, and the turning point of spherical shells in higher dimensions is located outside the horizon of outer spacetime, it is reasonable to speculate that the property of turning points for spherical shells' trajectory in high dimensions should be the same to that of Einstein case. To substantiate the argument, we provide a detailed analysis of the EGB case in five dimensions. 

In this subsection, we are interested in the trajectory of shell with $n_{o}^{r}>0$ in $r>r_{H}$. To determine the location of the turning point of the shell's trajectory, which is given by 
\[
r^{2}\left(\left(1-\sqrt{f_{+}}\right)+\frac{2}{3r^{2}}\left(\sqrt{f_{+}}(-3+f_{+})+2\right)\right)=m,
\]
is equivalent to find the zero points of     
\begin{eqnarray}
	G(r)=r^{2}\left(1-\sqrt{f_{+}}\right)+\frac{2}{3}\left(\sqrt{f_{+}}\left(-3+f_{+}\right)+2\right)-m, \nonumber
\end{eqnarray}
whose derived function is given by
\begin{eqnarray}
	G'(r)&=&2r\left(1-\sqrt{f_{+}}\right) \nonumber\\
	&&+\frac{1}{\sqrt{f_{+}}r}\left((f_{+}-1)(2(f_{+}-1)-r^{2})-\frac{8}{3}M\right),\nonumber
\end{eqnarray}
and asymptotic behavior at infinity is
\[
G(r\rightarrow+\8)=\left(-m+\frac{2}{3}M\right)+\frac{2}{9}M^{2}r^{-2}+O\left(r^{-3}\right).
\]

By analyzing the derivative function, we can see that $G(r)$ decreases monotonically in the interval $r>r_{H}$, indicating that a zero point for $G(r)$ exists in this interval if and only if the mass parameters satisfy 
\begin{eqnarray}
	\{m,M\}&=&\{G(r_{H})>0 \cap G(+\8)<0\} \nonumber \\
	&=&\{\frac{2}{3}M<m\leq\frac{1}{3}(1+4M)\}.\label{eq:27-1}
\end{eqnarray} However, we cannot conclusively determine which type of turning point exists in this case, is it an expanding shell with mass parameters (\ref{eq:27-1}), which is unable to escape its own gravitational pull, or is it a collapsing shell that bounces back to infinity? Nonetheless, we can numerically identify the type of turning point for certain mass parameters. Transforming (\ref{eq:16}) in five dimensions and obtain 
\begin{eqnarray}
	&&\dot{r}^{6}+\left(\frac{3}{2}r^{2}+3-\frac{4M^{2}}{9m^{2}}\right)\dot{r}^{4}+\left(\frac{3}{4}r^{2}+1\right)^{2}-\frac{9Q^{2}}{64m^{2}} \nonumber\\
	&&+\left(\left(\frac{3}{4}r^{2}+2\right)^{2}-1-\frac{MQ}{2m^{2}}\right)\dot{r}^{2}=0,\label{eq:28}
\end{eqnarray} 
where
\begin{eqnarray}
	Q=&&m^{2}+\frac{1}{18}\left(-1+\sqrt{1+\frac{16M}{3r^{4}}}\right)r^{6}\nonumber \\ &&+\frac{8}{27}M\left(6+\left(3+\sqrt{1+\frac{16M}{3r^{4}}}\right)r^{2}\right). \nonumber
\end{eqnarray} 

As with the four-dimensional case,  there exists a unique real solution for
(\ref{eq:28}), denoted by $\dot{r}^{2}=U(r,m,M)=V(r,\frac{m}{r},\frac{2M}{3r})$, where $U(r,m,M)\geq0$ corresponds to the physically allowable region of the spherical shell
trajectory. Without loss of universality, we take $M=5$ with different $m$ which satisfy (\ref{eq:27-1}) to solve $U(r)=0$ to determine the location of  the spherical shell's turning point, denoted by $r_{tp}$, and study the sign of $U'(r_{tp})$ to see the force at this position, see Figure 4. As illustrated in Figure 4, the spherical shell's force direction at these turning points is pointing to decreasing $r$, implying that the spherical shell with these mass parameters cannot escape its own gravitational pull.

\begin{figure}
	\includegraphics[scale=0.66]{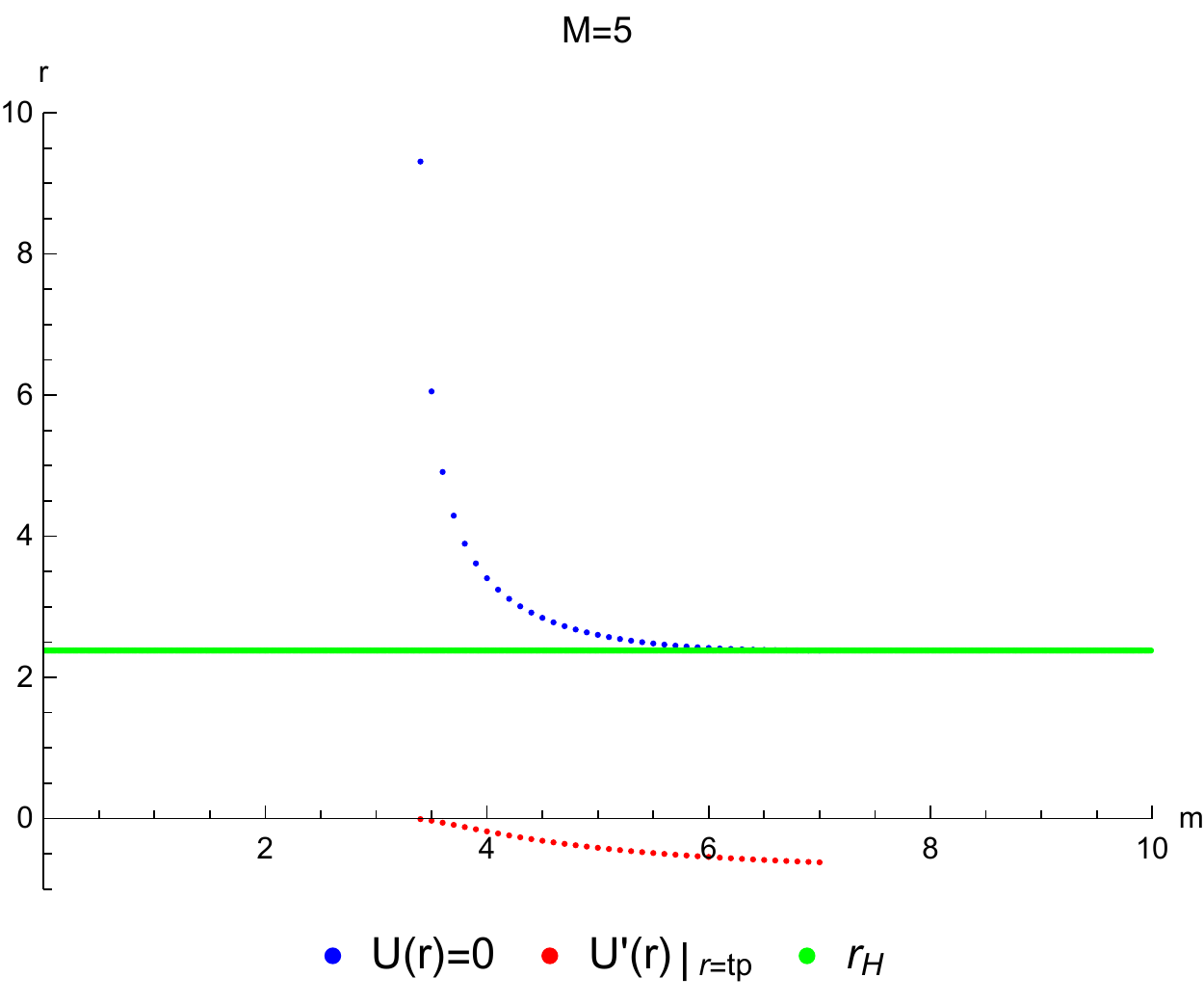}\caption{The location of turning points for $M=5$ with $m$ from 3.4 to 7 with $\protect\D m=0.1$ is plotted. Note that $U'(r)|_{r=tp}$ only has numerical meaning in the vertical direction.}
\end{figure}
        
        In summary, collapsing thin shells
	exhibit novel bouncing behavior in four dimensions, but not in higher dimensions, as confirmed by the detailed analysis in five dimensions. The bouncing behavior is indeed caused by the $\a$ term in four dimensions, whose effect is similar to that of electrical charge somehow, despite the fact that it contributes to Einstein's theory as a pure gravitational correction.	
\section{Weak cosmic censorship for thin shells}
The cosmic censorship has been tested by thin shells for many years. As we know, there is always a horizon for $M>0$ in Schwarzchild spacetime, however in 4D EGB gravity, we can deduce from equation (\ref{eq:1-2}) that only $M\geq\a$ makes the appearance of horizon, whereas $M<\a$, which we refer to an underweight spacetime, later describes spacetime with naked singularity. Thus, one might reasonably ask whether a shell with an underweight exterior $(M<\a)$ can implode past the horizons of an existing non-extremal black hole $(M>\a)$. If this occurs, a naked singularity forms and weak cosmic censorship is violated. Now, we're going to construct this scenario to test the cosmic censorship conjecture.

In this section, we are concentrated on the asymptotically flat region, where $n^{r}>0$ is always true. For our purposes, we will consider such a gravitational collapse process for a spherical shell, whose interior and exterior geometry are given by the spherical symmetric solution (\ref{eq:1-1}). Let $M_{i}$ and $M_{o}$ denote the interior and exterior mass parameters, respectively, we will have
\begin{eqnarray}
	F_{i}(r)&=&1+\frac{r^{2}}{2{\text{\ensuremath{\a}}}}\left(1-\sqrt{1+\frac{8{\a}M_{i}}{r^{3}}}\right),\nonumber \\ F_{o}(r)&=&1+\frac{r^{2}}{2{\text{\ensuremath{\a}}}}\left(1-\sqrt{1+\frac{8{\a}M_{o}}{r^{3}}}\right). \nonumber
\end{eqnarray} 
In this case, we specify $M_{i}>\a$ and $M_{o}<\a$. The motion of a spherical shell follows the equation (\ref{eq:4-4}), which we have obtained previously. We transform (\ref{eq:4-4}) in four dimensions and obtain
\begin{eqnarray}
	r(n_{i}^{r}-n_{o}^{r})&&\left(1+\frac{2{\a}}{3r^{2}}\left(3(1+\dot{r}{}^{2})-\left((n_{o}^{r}){}^{2}+n_{o}^{r}n_{i}^{r}+(n_{i}^{r})^{2}\right)\right)\right) \nonumber \\
	&&=m,\label{eq:29}
\end{eqnarray}
where\[n_{i}^{r}=\sqrt{F_{i}(r)+\dot{r}^{2}},\qquad n_{o}^{r}=\sqrt{F_{o}(r)+\dot{r}^{2}}.\]  Since $F(r)$ decreases monotonically from $M=0$ to $M=+\infty$, we have $F_{o}(r)>F_{i}(r)$ and $n_{o}^{r}>n_{i}^{r}$.
Due to the facts $0<n_{i}^{r}<n_{o}^{r}<\sqrt{1+\dot{r}^{2}}$, the left-hand side of (\ref{eq:29}) has
\begin{eqnarray}
	&&r(n_{i}^{r}-n_{o}^{r})\left(1+\frac{2{\a}}{3r^{2}}\left(3(1+\dot{r}{}^{2})-\left((n_{o}^{r}){}^{2}+n_{o}^{r}n_{i}^{r}+(n_{i}^{r})^{2}\right)\right)\right)\nonumber \\
	&&<r(n_{i}^{r}-n_{o}^{r})<0,
\end{eqnarray} 
which indicates that $m$ in (\ref{eq:29}) should be negative. However, because the weak energy condition in (\ref{eq:9-3}) implies that $\s(\t)>0$, which is identified with $m>0$, the hypothetical process is unphysical. In other words, it cannot violate the cosmic censorship via thin shells' dynamics in 4D EGB gravity, implying that the weak cosmic censorship conjecture holds. Moreover, the preceding analysis also can be used to test the third law of black hole dynamics, which states that no extremal black hole forms. The only modification required for the preceding discussion is to replace the outer spacetime with $M=\a$. In that case, the third law still applies.

\section{Concluding remarks}
In this paper, we have considered the gravitational collapse of a spherical dust thin shell in EGB gravity. The EOM for the thin shell in $D$ dimensions has been derived. Based on 4D EGB gravity, we are motivated by the trajectory of a collapsing dust star and discover novel bouncing behavior for the thin shell at small scale.
It is worth noting that the bouncing process is consistent with the
fact that the higher curvature corrections to GR modifies the small scale property of Einstein's gravity. The EOM is analyzed in four dimensions and  the oscillating shell is classified using the Penrose diagram. The analysis of EOM in higher dimensions is also included, where it is found that the collapsing thin shell cannot be bounced back (i.e. the singularity
always forms). Finally, we teste the weak cosmic censorship conjecture and find that it still holds in our model. However, some questions remain unanswered.
Is the oscillation behavior still true for more realistic cases, such
as thick shells, shells with more realistic equations of state and general (non-)spherical stars? Can dissipation
during the collapse be taken into account? What will happen then?
These questions will be left for future works.

\section*{Acknowledgments}
We are grateful to Hongbao Zhang for his helpful discussions. This work is partially
supported by NSFC with Grant No.11975235, 12035016, 11731001 and 11575286.

\section*{Appendix: the Penrose diagram of 4D EGB black hole}
The Penrose diagram is a crucial tool for analyzing spacetime in GR. We now construct the Penrose diagram of the spherically symmetric vacuum solution for 4D EGB gravity (\ref{eq:1-1}).

In this section, we are interested in the non-extremal case. In this case, the metric (\ref{eq:1-1}) is singular when $r=0, r_{\pm}$. Cutting the spacetime manifold into three disconnected regions in the coordinates $\{t,r,\theta,\phi\}$, then the regions are $0<r<r_{-}$, $r_{-}<r<r_{+}$ and $r_{+}<r<+\infty$, respectively. Given that we are considering the connected spacetime manifold, we initially choose the region $r_{+}<r<+\infty$, later named region $I_{+}$, to represent the external field. It can be showed the region is extensible. 

Let's introduce null coordinates $U,V$ in region $I_{+}$, which denotes
\[V = \exp (t+r_{*}) ,\qquad U = - \exp (r_{*}-t),\]
where the tortoise coordinate $r_{*}$ is defined by $r_{*}=\int dr F(r)^{-1}.$ Note that we have
\begin{eqnarray}
	&&{\lim_{r\to +\infty}}r_{*}=+\infty,\qquad{\lim_{r\to r_{+}}}r_{*}=-\infty,\nonumber \\
	&& {\lim_{r\to r_{-}}}r_{*}=+\infty, \qquad {\lim_{r\to 0}}r_{*}=c, \nonumber 
\end{eqnarray} where c is a constant. 

Using the coordinates $\{U,V,\theta,\phi\}$, the metric (\ref{eq:1-1}) takes the form
\begin{equation}
	ds^{2}=F(r) \exp (-2r_{*})dUdV+r^{2}\left(d\theta^{2}+\sin^{2}\theta   d\phi^{2}\right).\label{eq:36}
\end{equation}
One can prove $(\ref{eq:36})$ is no longer singular at $r=r_{+}$. This indicates the singularity $r=r_{+}$ is the result of a bad choice of coordinates, rather than a real singularity.

Next we use the arctangent to bring $U,V$ into a finite coordinate value and define
\begin{equation}
	V'=\arctan V,\qquad U'=\arctan U.\label{eq:37}
\end{equation}
For region $I_{+}$, its corresponding ranges given by
$$0<V'<\frac{\pi}{2},\qquad -\frac{\pi}{2}<U'<0,$$
and because $r=r_{+}$ is non-singular, it is naturally extend the region $I_{+}$ across $r=r_{+}$ to a new region which is isometric to the region $r_{-}<r<r_{+}$ of the 4D EGB solution (\ref{eq:1-1}), either along the direction of $\partial_{U'}$ or $\partial_{V'}$. For the direction of $\partial_{U'}$, the new region called region $II_{+}$ is parameterized  by
$$0<V'<\frac{\pi}{2};\qquad 0<U'<\frac{\pi}{2}.$$
Together with $r=r_{+}$, region $I_{+}$ and $II_{+}$, we obtain Penrose diagram of the region $r_{-}<r<+\infty$ for (\ref{eq:1-1}).

\begin{figure}
	\includegraphics[scale=0.72]{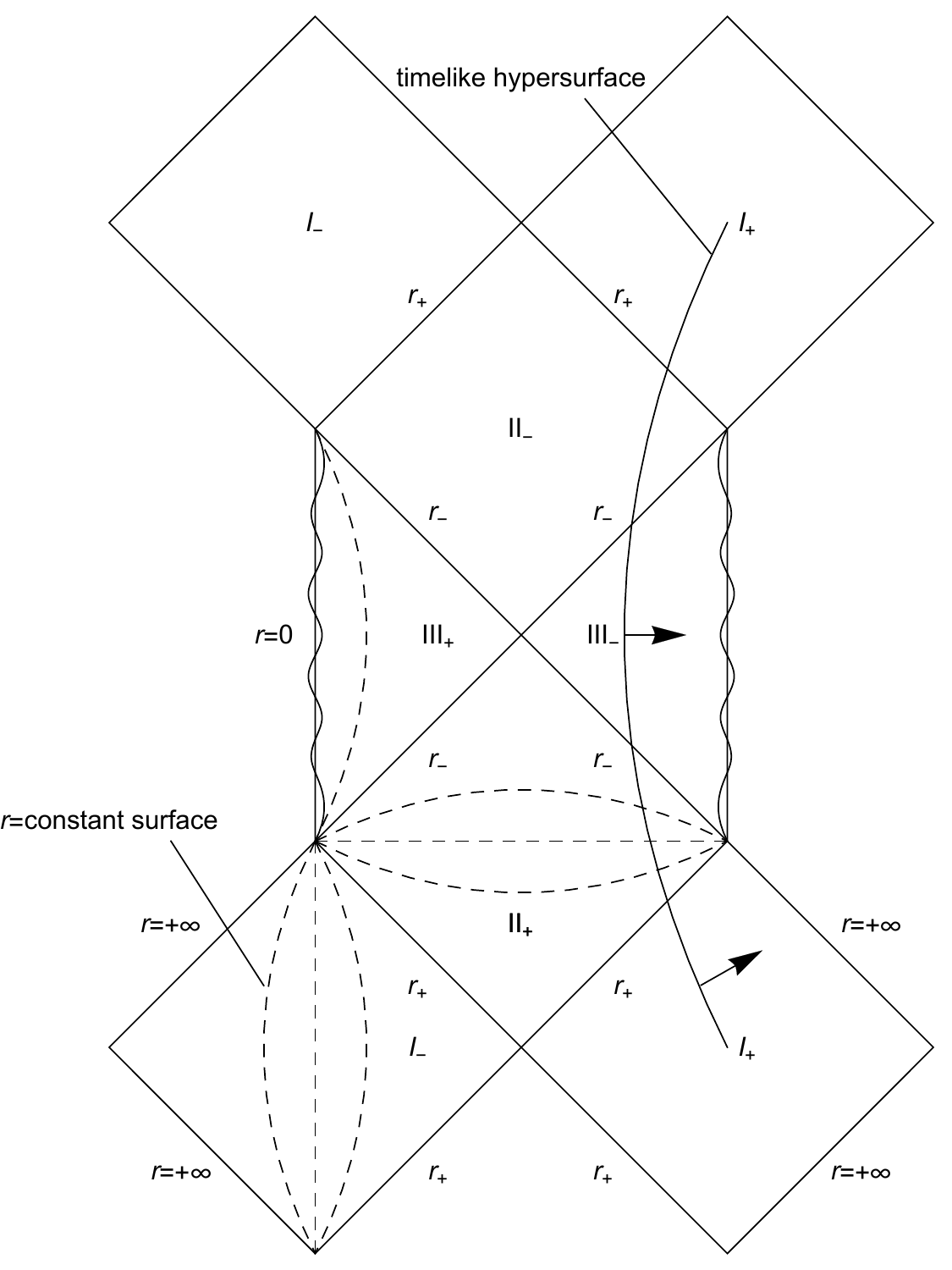}\caption{The Penrose diagram of a four-dimensional EGB vacuum solution with $M>\a$ is plotted.}
\end{figure}

Similarly, one can find that $r_{-}$ is also not a real singularity, allowing one to extend region $r_{-}<r<r_{+}$ to region $0<r<r_{-}$, which we labeled as region $III_{+}$. Extend the diagram in the direction $\partial_{U}$ or $\partial_{V}$ indefinitely, as in the RN case, we would obtain the extended Penrose diagram of 4D EGB solution, see Figure 5.

Finally, we introduce a timelike hypersurface to illustrate the meaning of the sign of $n_{o}^{r}$ on the Penrose diagram. According to our convention, its normal vector $n_{o}^a$ points from inside to outside, as illustrated by the arrow in Figure 5. The sign of $n_{o}^{r}$ depends on the direction of $n_{o}^{a}$ pointing. If $n_{o}^{a}$ points to larger $r$, then we have $n_{o}^{r}>0$, while $n_{o}^{a}$ points to smaller $r$ means $n_{o}^{r}<0$. Notice that the sign of $n_{o}^{r}$ varies from region to region. For instance, in region $I_{+}$, $n_{o}^{a}$ always points to larger $r$, indicating that $n_{o}^{r}>0$, whereas in region $III_{-}$, $n_{o}^{a}$ always points to smaller $r$, implying that $n_{o}^{r}<0$.


%
%

\end{document}